\documentclass[aps,preprint,pra]{revtex4}

\usepackage{latexsym}
\usepackage{amsmath}
\usepackage{graphicx}

\newcommand{\be}{\begin{equation}}
\newcommand{\ee}{\end{equation}}

\begin{document}

\title{Spatially Inhomogeneous Superconducting and Bosonic Networks with Emergent Complex Behaviors}

\author{F. P. Mancini $^1$, P. Sodano $^1$ and A. Trombettoni $^2$}
\affiliation{$^1$ Dipartimento di Fisica and Sezione
I.N.F.N., Universit\`a di Perugia, Via A. Pascoli, Perugia,
I-06123, Italy}
\affiliation{$^2$ S.I.S.S.A. and Sezione I.N.F.N., Via Beirut 2/4, I-34014
Trieste, Italy}

\begin{abstract}
The spontaneous emergence of enhanced responses and local orders are properties often associated with complex 
matter where nonlinearities and spatial inhomogeneities dominate. We discuss these phenomena 
in quantum devices realized with superconducting Josephson junction networks and cold atoms in optical lattices. We evidence how 
the pertinent engineering of the network's shape induces the enhancement of the zero-voltage Josephson critical currents 
in superconducting arrays as well as the emergence of spatially localized condensates for cold atoms in inhomogeneous 
optical lattices. \\
DFUPG: 39-07
\end{abstract}

\date{\today}
\maketitle

\section{Introduction}

In all fields of physics homogeneous systems have the simplest properties and, thus, play a very important paradigmatic role in our understanding of natural phenomena.
It is a fact, however, that many real systems are inhomogeneous in a way or another and that, in some instances, their inhomogeneity may be the seed for the emergence 
of new and unexpected complex behaviors \cite{complex}, which may be probed in experiments and, hopefully, become useful in the engineering of quantum devices.

In condensed matter systems, inhomogeneities may lead to enhanced responses to external perturbations and/or to the
emergence of local orders \cite{enhare}. Remarkable examples include the large transport anisotropy observed at
low-temperature in quantum Hall samples after the onset of electronic nematic phases \cite{nematic}, the colossal
magneto-resistance in manganites \cite{manganite}, the appearance of striped phases in systems with competing interactions
\cite{stripe} as well as of pseudogap phases in high $T_c$ superconductors \cite{pseudogap}. 
In all the above mentioned examples the relevant optimal inhomogeneities are dynamically generated \cite{kivefra}; 
furthermore, stripe and pseudogap phases are associated to the onset of local orders \cite{stripe,Ovchinnikov}.

To get control on the onset of complexity of a condensed matter system is desirable not only for understanding new 
emergent functionalities useful for the engineering of new materials and devices but also for discovering levels of theoretical
description enabling to well separate the properties of global phases from the ones arising from phase competition;
global average behaviors are, in fact, not helpful for this task \cite{anderson}. Quantum devices provide a controllable
setting (one can fabricate them!) to investigate the effects induced by inhomogeneities on the emergence of complex
behaviors.

Quantum devices with built-in inhomogeneities may be realized with today's available technologies using either
superconducting Josephson junction networks (JJN) \cite{moshe} or ultracold atoms in optical
lattices \cite{morsch06}. Due to their versatility and to the great reliability of the fabrication technologies
developed for their construction, JJNs and ultracold atoms in optical lattices 
are by now the prototype of complex physical systems exhibiting a
variety of interesting physical behaviors, adjustable by acting only on very few external parameters and, as we shall point out,
also by the pertinent engineering of the network's shape; in  addition, they provide controllable settings to investigate
the properties of granular superconductors or high $T_c$ superconductors \cite{simanek94} paving a very promising avenue
in the engineering of quantum states of potential interest for processing quantum information \cite{MSS}.

\begin{figure}[t]
\begin{center}
\includegraphics[scale=.55]{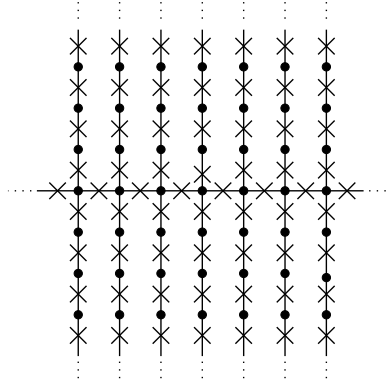}
\caption{\label{fig1} The comb graph.}
\end{center}
\vspace{8mm}
\end{figure}

A Josephson junction may be realized using two superconducting grains separated by an insulating layer; by
an appropriate engineering of the insulating support, one can fabricate an inhomogeneous network of Josephson junctions
with a given shape.
Inhomogeneous JJNs \cite{inhom,deut} have been studied for a long time with the aim of establishing the effect of space 
connectivity on superconductivity \cite{consup}. Recently, the appealing
perspective to realize devices for the manipulation of quantum information stimulated the analysis of inhomogeneous
planar JJNs with non conventional connectivity \cite{ioffe}, engineered to sustain a topologically ordered ground
state \cite{wen}. Furthermore, transport measurements on superconducting wire networks evidenced - even in pure systems
with non-dispersive eigenstates - interesting anomalies of the network critical current induced by the interplay between
the network's geometry and topology with externally applied magnetic fields \cite{abilio}. Furthermore, the
theoretical analysis of rhombi chains has evidenced the exciting possibility of being able to detect $4e$
superconductivity through measurements of the supercurrent in presence of a pertinent external
magnetic field \cite{feigel}. Here we shall address the properties of inhomogeneous JJNs fabricated on lattices with
non-random regular inhomogeneities engineered to yield enhanced zero-voltage Josephson critical currents as well as
local order on domains selected by the network's topology. We shall investigate in detail the paradigmatic example of
a comb-shaped JJN, whose properties have been analyzed in \cite{silvestrini05,sodano06}. To fix the ideas, in
Fig. \ref{fig1} is reported the design of the device; there, the circles locate the position of the superconducting grains
while the crosses represent the junction joining them.

Control over the network connectivity may be achieved also with ultracold atoms on optical lattices \cite{pitaevskii03}. 
A one-dimensional optical lattice is realized using two counter-propagating laser beams arranged to create a periodic potential where bosons may tunnel
between wells with a rate adjustable by tuning the laser's power controlling the height of the inter-well barrier. In
this realization a lattice site is a minimum of the periodic potential while the links between neighbor vertices are
provided by the barrier between wells. By suitably arranging more laser beams one is able to engineer the shape of the
optical networks \cite{oberthaler02} allowing, in principle, for an experimental testing of the proprieties
of inhomogeneous lattices such as the ones analyzed in \cite{roth03,brunelli04}. To fabricate bosonic Josephson
junctions (BJJ) and networks (BJJN) one should require also that all the atoms in a given well are described
by the same macroscopic wave function: a BJJ is obtained, then, by loading - at a temperature below the Bose-Einstein
condensation critical temperature - a condensate in each well of a double well potential \cite{oberthaler05} while
the barrier separating the two condensates acts as a Josephson link \cite{smerzi97}. With a
multi-well optical potential, provided that the heights of the barriers are much higher than the condensate's chemical
potential, one fabricates a BJJN (i.e., a lattice of weakly coupled condensates). In the following we shall evidence how
bosons hopping on comb-shaped optical lattices undergo a spatial Bose-Einstein condensation
on the comb's backbone; this happens even if the bosons are free and the network' s euclidean dimension is $1$.

The emergence of complex behaviors in the superconducting and bosonic arrays analyzed in this paper is determined solely by the
pertinent engineering of the network's connectivity, since it is due only to the spectral properties of the adjacency matrix
(i.e., of the matrix defined as $A_{ij}=1$ if $i$ and $j$ are vertices connected by a link, and $A_{ij}=0$
otherwise \cite{graf}), which fully characterizes the network's geometry and topology. In some very specific instances \cite{burioni00,burioni01}, 
the spectrum of $A_{ij}$ contains a continuous set of states - {\em the hidden spectrum} - localized around domains selected by the network's topology with
eigenvalues ranging {\em continuously} from a certain value $E_0$ up to the threshold of the continuum delocalized states $E_d$:
the spectrum is then effectively gapless even if, in the thermodynamic limit, there is a lowest eigenvalue $E_0$, confined
away from $E_d$. In the following we shall evidence that it is this spectral anomaly - induced from the pertinent choice of the network's connectivity-  which is 
responsible for the emergence of enhanced responses and local orderings in superconducting and bosonic 
networks, respectively.

An hidden spectrum of the adjacency matrix emerges, for instance, when one analyzes bundled graphs \cite{graf}
(i.e., those obtained by grafting a fiber graph to every point of a {\em base} graph) while, for graphs with constant
coordination number (such as the Sierpinski gasket and the ladder graph), the adjacency matrix does not support any hidden
spectrum \cite{burioni01}. In the following, we shall analyze the simple paradigmatic case of comb networks showing how the
hidden spectrum of the adjacency matrix leads to unusual quantum behaviors such as the emergence of the spatial BEC 
on the comb's backbone for a Bose gas living on a comb-shaped optical lattice \cite{burioni00,giusiano04} and of the enhanced responses observed for
classical comb-shaped JJNs made of Niobium grains \cite{silvestrini05,sodano06}[see Fig.1]. To better clarify our arguments,
we find instructive to compare our results with those obtainable if the same devices were defined on a chain,
since the latter is, after all, the simplest graph of euclidean dimension $1$.

In section 2 we determine the equations governing the properties of inhomogeneous superconducting and bosonic networks.  
Section 3 is devoted to the analysis of the spectrum of the adjacency matrix of a comb-shaped network and to a comparison of its spectrum with the one pertinent
to a linear chain: there we see that, while the adjacency matrices of both graphs admit a continuum set of delocalized states starting at
the same eigenvalue, the adjacency matrix of a comb supports an hidden spectrum. In section 4 we evidence how the states belonging
to the hidden spectrum are responsible for the enhancement of the zero-voltage Josephson critical currents in superconducting JJNs as
well as for the emergence of the spatial BEC in bosonic networks on domains pertinently selected by the network's topology. Section 5 is
devoted to our conclusions and final remarks while the appendices report on some pertinent but rather lengthy computational details.

\section{Many-body theory of superconducting and bosonic systems on generic networks}

In this section we shall summarize the many-body description of superconducting and bosonic networks. Our purpose here is mainly to clarify the assumptions 
underlying the equations used in section 4 to characterize the complex behaviors emerging from a pertinent choice of the connectivity in JJNs and bosonic networks.
Our approach uses the self- consistent Bogoliubov-De Gennes (BDG) equations \cite{degennes}  since they provide an unified framework to account for the 
description of fermionic and bosonic systems enabling to appreciate in a rather simple context the differences (and similarities) between these systems.

We shall describe first the microscopic theory of inhomogeneous superconducting networks realized by putting on each site of a graph a superconducting grain.
 On each island, the effects of the electron-phonon and Coulomb interactions are embodied in the BCS parameter and, furthermore, there exist a critical temperature $T_{BCS}$ 
such that for $T$ lesser than $T_{BCS}$ each grain becomes superconducting; of course, the entire network becomes superconducting at a lower temperature $T_c$ \cite{abeles}. 
We, then, analyze ultracold atoms (spinless, for simplicity) in deep optical lattices and derive the set of equations describing both the excitations
and the condensate's dynamics; 
in dealing with atomic systems one should only account for repulsive interactions between bosons as it is well known that attractive interactions lead to instabilities 
even in presence of confining traps \cite{pitaevskii03}.

\subsection{Superconducting networks}

Inhomogeneous fermionic systems with attractive interactions may be conveniently described by the  BdG equations
\cite{degennes}: with a two-body point-like interaction
$V(\vec{r}-\vec{r}')= -{\cal V} \delta(\vec{r}-\vec{r}')$, the
Hamiltonian is ${\cal H}={\cal H}_0+ {\cal H}_1$, where
\begin{equation}
{\cal H}_0=\int d\vec{r} \sum_{\sigma} \Psi^{\dag}(\vec{r} \sigma) H_0
\Psi(\vec{r} \sigma)
\label{H0}
\end{equation}
and
\begin{equation}
{\cal H}_1=-\frac{{\cal V}}{2}
\int d\vec{r} \sum_{\sigma \sigma'}
\Psi^{\dag}(\vec{r} \sigma) \Psi^{\dag}(\vec{r} \sigma')
\Psi(\vec{r} \sigma') \Psi(\vec{r} \sigma).
\label{H1}
\end{equation}
In Eqs.~(\ref{H0}) and (\ref{H1}), the $\Psi$'s are fermionic operators, $\sigma=\pm$ is a spin index, and $H_0=-\hbar^2
\nabla^2/2m+U_0(\vec{r})-\mu$, where $\mu$ is the chemical potential and $U_0(\vec{r})$ is an external potential
(we assume, as usual, that
$U_0(\vec{r})$ is spin-independent and that no magnetic field is
applied). When the fermions are not interacting, then 
$\mu=E_F$, where $E_F$ is the Fermi energy. In a self-consistent approach, one defines the effective
Hamiltonian
$${\cal H}_{eff}=\int d\vec{r}
\sum_{\sigma} \Big\{ \Psi^{\dag}(\vec{r} \sigma) H_0
\Psi(\vec{r} \sigma) + \sum_{\sigma} U(\vec{r}) \Psi^{\dag}(\vec{r} \sigma)
\Psi(\vec{r} \sigma) \Big\} $$
\begin{equation}
+\int d\vec{r} \Big\{ \Delta(\vec{r}) \Psi^{\dag}(\vec{r} +)
\Psi^{\dag} (\vec{r} -) + \Delta^{\ast} (\vec{r}) \Psi(\vec{r} -)
\Psi(\vec{r} +) \Big\}
\label{Heff}
\end{equation}
and, then, requires that the {\em pair potential} $\Delta(\vec{r}) \equiv {\cal V} \langle 
\Psi(\vec{r}+) \Psi(\vec{r}-)\rangle$
and the {\em Hartree-Fock potential} $U(\vec{r}) \equiv - {\cal V} \langle 
\Psi^{\dag} (\vec{r} \sigma) \Psi(\vec{r} \sigma) \rangle$ are self-consistently
determined from the solutions of the BdG
equations
\begin{equation}
\in_\alpha u_\alpha(\vec{r}) = [ H_0 + U(\vec{r}) ] u_\alpha(\vec{r}) +
\Delta(\vec{r}) v_\alpha(\vec{r})
\label{BdG1}
\end{equation}
\begin{equation}
\in_\alpha v_\alpha(\vec{r}) = - [ H_0 + U(\vec{r}) ]
v_\alpha(\vec{r}) + \Delta^{\ast}(\vec{r}) u_\alpha(\vec{r}),
\label{BdG2}
\end{equation}
with $u_{\alpha}$, $v_{\alpha}$ satisfying 
\begin{equation}
\int d\vec{r} \left[
\vert u_{\alpha}(\vec{r})\vert^2 + \vert v_{\alpha}(\vec{r})\vert^2
\right]=1.\label{norm1}
\end{equation} 
One has then
\begin{equation}
\Delta(\vec{r})={\cal V}
\sum_\alpha  u_\alpha(\vec{r})
v^{\ast}_{\alpha}(\vec{r}) \tanh{\left( \frac{\beta}{2}
\in_\alpha \right)}
\label{Delta}
\end{equation}
and
\begin{equation}
U(\vec{r})=-{\cal V}
\sum_\alpha \left[ \vert u_\alpha(\vec{r}) \vert^2 f_\alpha +
\vert v_\alpha(\vec{r}) \vert^2 \left( 1-f_\alpha \right) \right],
\label{U}
\end{equation}
where $f_\alpha=(e^{\beta \in_\alpha}+1)^{-1}$, $\beta=1/k_B T$ and
the sums are only taken on positive $\in_\alpha$'s. The chemical potential $\mu$ - when different 
from the Fermi energy - is determined 
from the normalization condition. Of course, when
there is an external potential breaking the translational invariance,
the pair potential is dependent on the position.

For spatially homogeneous superconducting networks the quantum number $\alpha$ is
just the momentum $\vec{k}$ and, then, $\Delta= \Delta(\vec{r})$ and
$U=U(\vec{r})$ do not depend on the position implying that
$\in_{\vec{k}}=\sqrt{\Delta^2+E_{\vec{k}}^2}$, where
$E_{\vec{k}}=\hbar^2 k^2/2m-\mu+U$. Furthermore, since
$u_{\vec{k}}(\vec{r})=L^{-3/2} U_{\vec{k}} e^{i {\vec{k}} \cdot
{\vec{r}}}$ and $v_{\vec{k}}(\vec{r})=L^{-3/2} V_{\vec{k}} e^{i
{\vec{k}} \cdot {\vec{r}}}$ with $L^3$ being the volume, one gets
$U_{\vec{k}}^2=(1/2) [1+E_{\vec{k}}/\in_{\vec{k}}]$ and
$V_{\vec{k}}^2=(1/2) [1-E_{\vec{k}}/\in_{\vec{k}}]$. With the 
BCS interaction (i.e., ${\cal V}=V_{BCS}$ if 
$\vert \hbar^2 k^2/2m-E_F \vert$, $\vert \hbar^2 k'^2/2m-E_F \vert < \hbar \omega_D$ and ${\cal V}=0$
otherwise, $\omega_D$ being the Debye frequency) and $U=0$
(which is a good solution of the self-consistent equation for $U$ if
$n(0)V_{BCS} \ll 1$), one has $\mu=E_F$ together with the celebrated BCS gap-equation 
\begin{equation}
1=\frac{n(0)V_{BCS}}{2}
\int_{-\hbar \omega_D}^{\hbar \omega_D} \frac{dE}{\sqrt{\Delta^2+E^2}}
\tanh{\left( \frac{\beta}{2}\sqrt{\Delta^2+E^2} \right)};
\label{gapBCS}
\end{equation}
$n(0)$ is the density of states per volume and 
spin direction at the Fermi energy.
For inhomogeneous networks, instead, one should regard the external potential 
$U_0(\vec{r})$ as representing the insulating barriers between the grains and take self-consistently into account its effects.

To obtain a discrete version of the BdG equations (LBdG) suitable to describe the classical superconducting JJNs
fabricated in \cite{silvestrini05}, one makes the ansatz that the eigenfunctions of the continuous BdG
equations \cite{degennes} may
be written in a tight binding form as $u_\alpha(\vec{r}) = \sum_i
u_\alpha(i) \phi_i(\vec{r})$ and $v_\alpha(\vec{r}) = \sum_i
v_\alpha(i) \phi_i(\vec{r})$; $i$ labels the position of a
superconducting island while the contribution of the electronic
states participating to superconductivity in a given island is
effectively described by a field $\phi_i(\vec{r})$, whose specific
form depends only on the geometry of the islands and on the
fabrication parameters of the connecting junctions. The assumption
that $\phi_i(\vec{r})$ does not depend on $\alpha$ amounts to
account only for contributions coming from electrons near the Fermi
surface. The LBdG equations then read
\begin{equation}
\in_\alpha u_\alpha(i) = \sum_j \epsilon_{ij} u_\alpha(j) +
\Delta(i) v_\alpha(i) \label{LBdG1}
\end{equation}
\begin{equation}
\in_\alpha v_\alpha(i) = - \sum_j \epsilon_{ij} v_\alpha(j) +
\Delta^{\ast}(i) u_\alpha(i). \label{LBdG2}
\end{equation}
where $u_\alpha$ and $v_\alpha$ satisfy to
\begin{equation}
\sum_i \left[ \vert u_{\alpha}(i)\vert^2 + \vert
v_{\alpha}(i)\vert^2 \right]=1 . \label{complete}
\end{equation}
The matrix $\epsilon_{ij}$ is defined by
\begin{equation}
\epsilon_{ij}=-t A_{ij}
+ U(i) \delta_{ij} - \tilde{\mu} \delta_{ij}, \label{epsilon}
\end{equation}
with $A_{ij}$ being the adjacency matrix of the network,
\begin{equation}
\tilde{\mu}=\mu-\int d\vec{r} \phi_i(\vec{r}) \left( -\hbar^2
\nabla^2 / 2m \right) \phi_i(\vec{r}) \label{mutilde}
\end{equation}
and
\begin{equation}
t = - \int d\vec{r}
\phi_i(\vec{r}) [ -\hbar^2 \nabla^2/2m + U_0(\vec{r})]
\phi_j(\vec{r}). \label{hophop}
\end{equation}
Self-consistency
requires
\begin{equation}
\Delta(i)=\tilde{\cal V} \sum_\alpha  u_\alpha(i)
v^{\ast}_{\alpha}(i) \tanh{\left( \frac{\beta}{2} \in_\alpha
\right)} \label{scdelta}
\end{equation}
and
\begin{equation}
U(i)=-\tilde{\cal V} \sum_\alpha \left[ \vert u_\alpha(i) \vert^2
f_\alpha + \vert v_\alpha(i) \vert ^2 \left( 1-f_\alpha \right)
\right], \label{scu}
\end{equation}
where $\tilde{\cal V} \equiv {\cal V} \phi^2(\vec{r}=\vec{r}_i)$ is
assumed to be independent of $i$. The network's topology and geometry is encoded in the term
$-t A_{ij}$ appearing in the definition of the matrix
$\epsilon_{ij}$ given in Eq.~(\ref{epsilon}), while the specific values of $t$ and $\tilde{\cal
V}$ depend - as a result of our ansatz on the form of the
eigenfunctions of the BdG equations- only on the $\phi_i(\vec{r})$.

To justify the assumptions involved in the derivation of
Eqs.~(\ref{LBdG1}) and (\ref{LBdG2}), one should observe that, for the JJN device
described in \cite{silvestrini05}, capacitive (inter islands and
with a ground) effects are negligible, that the total number of
electrons on the island ${\cal N}$ is much larger than the number of
electrons tunneling through the Josephson junction and that all the
islands contain approximately the same ${\cal N}$ (${\cal N}(i)
\equiv {\cal N}$). Furthermore, the islands are big enough to
support the same superconducting gap of the bulk material; as a
result, one may require $\phi_i(\vec{r})$ to be position-independent
on each island except for a small region near the junction and to be
the same on each island with a normalization given by $\int d\vec{r}
\phi_i^2(\vec{r})= {\cal N}(i) \equiv {\cal N}$ and
$\int d\vec{r} \phi_i(\vec{r}) \phi_j(\vec{r})\approx 0$ for $ i
\neq j$; in our derivations we set ${\cal N} \equiv 1$. 
As a result $t \approx E_J=(\hbar/2e)I_c$, where $E_J$ is the nominal
value of the Josephson energy of all the junctions in the network
while $I_c$ is the {\em bare} zero-voltage Josephson critical
current of each junction. In section IV
we shall provide an explicit solution of the BdG equations 
(\ref{LBdG1}) and (\ref{LBdG2}) describing superconducting JJNs fabricated on a comb-shaped insulating support 
\cite{silvestrini05}, 
while in App. A a solution of the LBdG equations for a Josephson junction chain is provided.

\subsection{Bosonic networks}
The full quantum Hamiltonian pertaining to a (spinless) bosonic gas in 
an optical potential is ${\cal H}={\cal H}_0+ {\cal H}_1$, where
\begin{equation}
{\cal H}_0=\int d\vec{r} \Phi^{\dag}(\vec{r}) H_0
\Phi(\vec{r})
\label{H0-B}
\end{equation}
and
\begin{equation}
{\cal H}_1=\frac{g_0}{2}
\int d\vec{r} \Phi^{\dag}(\vec{r}) \Phi^{\dag}(\vec{r})
\Phi(\vec{r}) \Phi(\vec{r}).
\label{H1-B}
\end{equation}
In Eqs.~(\ref{H0-B}) and (\ref{H1-B}), $\Phi$ is bosonic operator, and $H_0=-\hbar^2
\nabla^2/2m+U_0(\vec{r})-\mu$; $U_0(\vec{r})$ is the external potential, which is usually the sum of the optical 
lattice and of the magnetic trap potential; for simplicity, we do not consider terms accounting for the effects of 
harmonic traps. In writing Eqs.~(\ref{H0-B}) and (\ref{H1-B}), 
the standard $s$-wave scattering approximation has been used: 
i.e., the two-body potential is written as $V(\vec{r}-\vec{r'})=g_0\delta(\vec{r}-\vec{r'})$, where 
$g_0=4 \pi \hbar^2 a / m > 0$,
with $a$ being the 
$s$-wave scattering length and $m$ the atomic mass. 
The dynamics of the bosonic field is described by the well known equation \cite{pitaevskii03}
\begin{equation}
i \hbar \frac{\partial}{\partial t}\Phi = 
\Big[H_0+g_0 \Phi^{\dag} \Phi \big] \Phi
\label{HAM-t}
\end{equation}
from which the Gross-Pitaevskii equation 
for the condensate wavefunction $\psi=\langle \Phi \rangle$ is usually derived \cite{pitaevskii03}. 

The bosonic counterpart of the BdG Eqs.~(\ref{BdG1}) and (\ref{BdG2}) is provided by the Bogoliubov equations for the condensate's excitations \cite{pitaevskii03}: 
to derive them, one usually writes $\Phi=\psi+\delta \Phi$ with 
\begin{equation}
\delta \Phi = \sum_\alpha [u_\alpha(\vec{r}) a_\alpha e^{-i \in_\alpha t /\hbar} - 
v^\ast_\alpha(\vec{r}) a_\alpha^{\dag} e^{i \in_\alpha t/\hbar }], 
\label{bogo1}
\end{equation}
where $a_\alpha$ are operators destroying bosons in the excited state $\alpha$. Using 
Eq.~(\ref{HAM-t}) and keeping only terms linear in the fluctuation field 
$\delta \Phi$, one finds
\begin{equation}
\in_\alpha u_\alpha(\vec{r}) = [ H_0 + 2 g_0 n_0(\vec{r})] u_\alpha(\vec{r}) -
g_0 n_0(\vec{r}) v_\alpha(\vec{r})
\label{BdG1-B}
\end{equation}
\begin{equation}
\in_\alpha u_\alpha(\vec{r}) = [ - H_0 - 2 g_0 n_0(\vec{r})] u_\alpha(\vec{r}) +
g_0 n_0(\vec{r}) v_\alpha(\vec{r})
\label{BdG2-B}
\end{equation}
where $n_0(\vec{r}) = \vert \psi(\vec{r}) \vert^2$ is the condensate density \cite{pitaevskii03}. Notice that $u_\alpha$ and $v_\beta$ satisfy now to the condition 
\begin{equation}
\int d\vec{r} 
[\vert u_\alpha(\vec{r}) \vert^2 - \vert v_\alpha(\vec{r}) \vert^2]=1. \label{normo2} 
\end{equation}

In the analysis of a bosonic system one has to use, in addition to the BDG equations, the Gross-Pitaevskii equation 
for the condensate. 
To do this, one should observe that, when the power laser is high enough, the bosonic field $\Phi$ may be 
approximated by means of the 
tight-binding approximation \cite{jaksch98} as
\begin{equation}
\label{ord-par-quant}
\Phi(\vec{r},t)=\sum_j b_j(t) \phi_j(\vec{r});
\end{equation}
substituting this ansatz in the full Hamiltonian, one gets (in the non interacting limit) a simple 
tight-binding model described by
\begin{equation}
H= - t \sum_{ij} A_{ij} b^{\dag}_i b_{j}, 
\label{B-H}
\end{equation}
where the coefficient $t$ is given by Eq.~(\ref{hophop}). In Eq.~(\ref{B-H}), 
$i,j$ denote the minima of the optical lattice (i.e., sites of the network) and
$b^{\dag}_j$ ($b_j$) is the bosonic operator which
creates (destroys) a boson at site $j$. The filling, i.e., the average number of particles per site, is
defined as $f=N_T/N_S$, where $N_T$ is the total number of bosons
and $N_S$ is the total number of sites.

Eq.~(\ref{B-H}) is the pertinent equation to investigate in order to ascertain if , for a gas of ultracold bosons, 
BEC emerges as a result of the network's inhomogeneity.
When BEC occurs in each well, each pair of neighbouring wells acts as a bosonic junction with {\em Josephson energy} given by
\begin{equation}
E_J \approx 2 t f. \label{Jos-ij}
\end{equation} 

In the next sections we shall show that a condensate indeed emerges when bosons hop on comb-shaped networks.

\section{Spectrum of a quantum particle hopping on comb-shaped networks}

In section 2 we evidenced how the equations describing 
superconducting and bosonic networks depend on the adjacency matrix characterizing 
the network's connectivity. In this section we shall review \cite{burioni00} the main results concerning the spectrum
of the adjacency matrix describing the connectivity of a comb network.

A comb (see Fig.~\ref{fig1}) is made of one-dimensional chains ({\em
fingers}) grafted periodically on a linear chain ({\em backbone}).
Each site of the comb can be naturally labeled by introducing two
integer indices $(x, y)$, where $x=0,\cdots,L_1$ labels the
different fingers and $y=0,\cdots,L_2$ provides the distance from
the backbone. Each site on the finger is linked to two neighbors
whereas each site of the backbone has four neighbors.

The topology of the network is fully described by the adjacency
matrix $A_{x,y;\:x',y'}$ which equals 1 if $(x,y; \: x',y')$  is a
link and $0$ otherwise.  A quantum particle hopping on a comb is,
then, described by the Hamiltonian
\begin{equation}
H=-t\sum_{x,y;x',y'} A_{x,y;\:x',y'}\: b^{\dag}_{x,y}
b_{x',y'}. \label{pure_hopping}
\end{equation}
The single-particle energy spectrum is found by solving the
eigenvalue equation \cite{burioni00,burioni01,giusiano04}:
\begin{equation}
-t \sum_{x',y'} A_{x,y;\:x',y'}\: \: \psi_E(x',y')= E \psi_E(x,y);
 \label{eigen_eq}
\end{equation}
on a comb  $A_{x,y;\:x',y'}$ is  given by
\begin{equation}
A_{x,y;\:x',y'}=
(\delta_{x,x'+1}+\delta_{x,x'-1})\delta_{y,0}\delta_ { 0 , y ' } +
(\delta_{y,y'+1}+\delta_{y,y'-1})\delta_{x,x'}.
\label{adjacency_comb}
\end{equation}
In the  following, we shall determine the spectrum of a quantum
particle on the finite $L_1\times L_2$ comb and, only at the end,
take the limit $L_1, L_2 \to \infty $. On a finite $L_1\times L_2$
comb - using the adjacency matrix \eqref{adjacency_comb}
- the
eigenvalue equation (\ref{eigen_eq}) reads:
\begin{equation}
-t\sum_{x'=0}^{L_1-1}\sum_{y'=0}^{L_2-1}\left[(\delta_{x,x'+1}+\delta_{x,x'-1})\delta_{y
,0}\delta_{ 0 , y ' }
+(\delta_{y,y'+1}+\delta_{y,y'-1})\delta_{x,x'}\right] \psi(x',y')=E
\psi(x,y) . \label{ee_comb}
\end{equation}
Without loss of generality one may take  $x$ and $y$ to be positive
integers, since, due to periodic boundary conditions, $(0,0)\equiv (L_1,0)$ and $(x,0)\equiv (x,L_2)$. The
total number of sites is then $N_S=L_1\times L_2$. 

By exploiting the
translation invariance along the backbone, a Fourier transform in
the variable $x$ reduces Eq. \eqref{ee_comb} to a one-dimensional
eigenvalue problem for a quantum particle hopping on the comb's
fingers. In fact, upon defining
\begin{equation}
\psi(k,y)=\sum_{x} e^{ikx}\psi(x,y) , \label{ft1}
\end{equation}
with $k=2\pi n/L_1$, $n=0,\dots L_1-1$, the eigenvalue equation
\eqref{ee_comb} becomes:
\begin{equation}
-t\sum_{k'=0}^{L_1-1}\sum_{y'=0}^{L_2-1}
\left[2\cos(k)\delta_{y,0}\delta_{0,y'}\delta_{k,k'}+(\delta_{y,y'+1}+
\delta_{y,y'-1})\delta_{k,k'}\right]\psi(k',y')=E\psi(k,y).
 \label{ee2}
\end{equation}
Since Eq.~(\ref{ee2}) is diagonal in $k$, it may be
written as:
\begin{equation}
-t\sum_{y'}\left[2\cos(k_0)\delta_{y,0}\delta_{0,y'}+
(\delta_{y,y'+1}+\delta_{y,y'-1})\right] \psi(y')=E \psi(y),
\label{comb_egv_eq}
\end{equation}
where $\psi(k,y)=\delta(k-k_0)\psi(y)$ with $k_0=2\pi n/L_1$,
$n=0,\dots, L_1-1$. Equation (\ref{comb_egv_eq}) describes then a
one-dimensional quantum particle interacting with a potential
located on the backbone, $V(k_0)=-2t \cos(k_0)$.

To determine the eigenvalues and eigenvectors of Eq.
\eqref{comb_egv_eq} one may look for solutions of the form
\begin{eqnarray}
\psi(y)=A_0 \cos(hy+\alpha) & {\rm\ \ \ for\ \ \ } & -2t\leq E
=-2t\cos(h)\leq 2t;
 \label{sol1}
 \\
\psi(y)=A_- \, e^{-hy}+B_- \, e^{hy} & {\rm \ \ \ for\ \ \ } & E=-t
(e^h+e^{-h})<-2t; \label{sol2}
\\
\psi(y)=A_+ (-1)^ye^{-hy}+B_+(-1)^ye^{hy} & {\rm \ \ \ for\ \ \ } &
E=t (e^h+e^{-h})>2t. \label{sol3}
\end{eqnarray}
To fix both the free parameters and the eigenvalues $E$ one requires
$\psi(y)$ to be normalizable and to be a solution of the eigenvalue
equation in $y=0$ and $y=L_2-1$. These points are the only ones
where Eq. \eqref{comb_egv_eq} is not identically satisfied, yielding
two equations to determine the two free parameters.
Since for a given value of $k_0$ there are $L_2$ different
eigenvalues, the spectrum will consist of $L_1 \cdot L_2$ states and
it can be divided in three regions: $\sigma_0$ and $\sigma_\pm$ \cite{burioni00}.

\begin{itemize}
 \item $\sigma_0$
 \end{itemize}
$\sigma_0$  is the part of the spectrum corresponding to delocalized
states with energies between $-2t$ and $2t$. Requiring the
wavefunction \eqref{sol1} to be a solution of the eigenvalue
equation \eqref{comb_egv_eq} in $y=L_2-1$ and $y=0$ yields
\begin{equation}
\begin{split}
&\cos[h(L_2-2)+\alpha]+\cos(\alpha) =2\cos(h)\cos[h(L_2-1)+\alpha]
\\
&\cos[h(L_2-1)+\alpha]+\cos(h+\alpha)+2\cos(k_0)\cos(\alpha)=
2\cos(h)\cos(\alpha) ,
\end{split}
\label{ee_del}
\end{equation}
 implying that there are odd eigenfunctions with
$\alpha=\pi/2$, $h=2\pi m/L_2$ and $m=1,\dots,L_2/2-1$ and $L_2/2$
even solutions, obtained by
\begin{equation}
-\cos(k_0) \cot(hL_2/2)=\sin(h). \label{eq_ac}
\end{equation}
Equation \eqref{eq_ac} can be solved graphically. In the large $L_2$
limit the allowed values for $h$ are: $h \approx \pi(2m -1)/L_2$,
with $m=1,\dots,L_2/2$. For each value of $k_0$ there are $L_2-1$
eigenvalues of type \eqref{sol1} with energy $E=-2t\cos(h)$ and
wavefunctions $\psi(x,y)=e^{ik_0x}\sin(hy)$ and
$\psi(x,y)=e^{ik_0x}\cos(h|y|+\alpha)$, with $h$ and $\alpha$
satisfying Eq. \eqref{ee_del}. Thus, the fraction of states in this
spectral region is $f=L_1(L_2-1)/L_1 \cdot L_2 $. Of course, $f$ tends
to 1 in the limit $L_1$, $L_2 \to \infty$ and the density of states is given by \cite{burioni01}:
\begin{equation}
\rho_0(E) \, dE= dn=L_1 d(L_2 h/2\pi)=L_1 (L_2-1)\, \frac{dE}{\pi
\sqrt{4t^2-E^2}} , \label{rho_catena}
\end{equation}
just as for a particle hopping on a linear chain.


\begin{itemize}

\item $\sigma_-$
 \end{itemize}
$\sigma_-$  is the part of the spectrum corresponding to localized
states with energies $E<-2t$. Requiring that the wavefunction
\eqref{sol2} is a solution of the eigenvalue equation
\eqref{comb_egv_eq} in $y=L_2-1$ and $y=0$ yields now
\begin{equation}
\begin{split}
&A_{-}\,e^{-h(L_2-2)}+B_{-}\,e^{h(L_2-2)}+(A_{-}+B_{-})
=(e^h+e^{-h})(A_{-}\,e^{-h(L_2-1)}+B_{-}\,e^{h(L_2-1)})
\\
&A_{-}e^{-h(L_2-1)}+B_{-}e^{h(L_2-1)}+(A_{-}e^{-h}+B_{-}e^{h})+2\cos(k
_ 0)(A_{-}+B_{-}) =(e^h+e^{-h})(A_{-}+B_{-}),
 \end{split}
 \label{eq_del}
\end{equation}
leading to:
\begin{equation}
\cos(k_0) \coth(hL_2/2)=\sinh(h).
 \label{low_hs}
\end{equation}
Equation (\ref{low_hs}) can be solved graphically, yielding a real
solution only if $\cos(k_0)>0$. The density of states is then given by
\begin{equation}
\rho_- (E)\, dE= dn= d(L_1 k_0/2\pi)=L_1 \, \frac{\vert E \vert
dE}{2\pi \sqrt{8t^2-E^2} \sqrt{E^2- 4t^2}}. \label{rho_hs_inf}
\end{equation}

\begin{itemize}

\item $\sigma_+$
 \end{itemize}
$\sigma_+$  is the part of the spectrum corresponding to localized
states with energies $E>2t$. The parameters $A_+$ and $B_+$ are
fixed by requiring again that the wavefunction \eqref{sol3}  is a solution
of the eigenvalue equation \eqref{comb_egv_eq} in $y=L_2-1$ and
$y=0$. This requirement yields a set of equations similar to those
of Eq.  \eqref{low_hs}, leading to $\cos(k_0)
\coth(hL_2/2)=-\sinh(h)$, which supports a real solution
only if $\cos(k_0)<0$ while, for $\cos(k_0)=0$, one gets the constant
solution $\psi(y)=1$ with energy $E=0$. The density of states in this spectral
region is given by
\begin{equation}
\rho_+ (E) \,dE= dn= d(L_1 k_0/2\pi)=L_1 \, \frac{\vert E \vert
dE}{2\pi \sqrt{8t^2-E^2} \sqrt{E^2- 4t^2}}. \label{rho_hs_sup}
\end{equation}

\begin{figure}
\begin{center}
\includegraphics[scale=0.4]{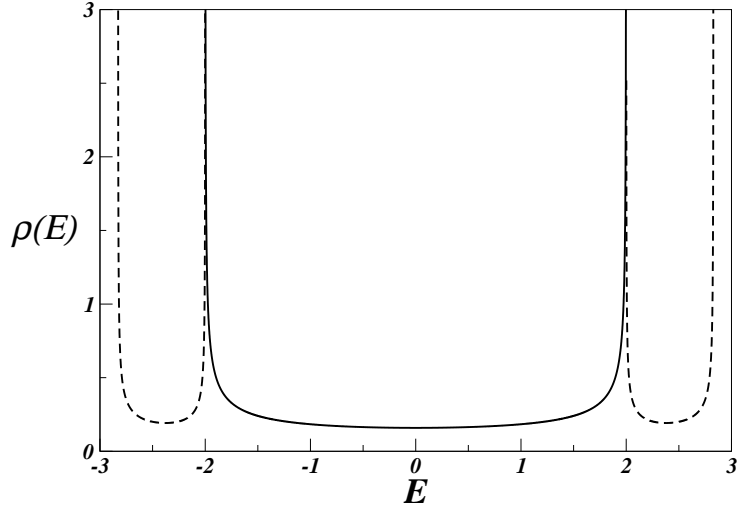}
\caption{\label{combo2} The density of states of the Hamiltonian
\eqref{pure_hopping} in units of $t $. The solid line indicates the
continuous part of the spectrum $\rho_0$, normalized to $L_1 \cdot
(L_2-1)$. The dashed lines denote  $\rho_\pm$, normalized to $L_1$.}
\end{center}
\vspace{8mm}
\end{figure}

The union of  $\sigma_{-}$ (i.e., states with $E<-2t$) and
$\sigma_{+}$ ($E>2t$) forms the hidden spectrum and
is the part of the spectrum corresponding to localized states; each spectral region contains $L_1/2$ states and, thus, $L_1$ states belong to the hidden spectrum. 
Taking the limit $L_2$ and $L_1 \to \infty$, from Eq. \eqref{low_hs}, one easily verifies that the states of the
hidden spectrum satisfy $\sinh(h)=\cos(k_0)$ and correspond to
energy eigenvalues $E=-2t \sqrt{1+cos^2(k_0)}$ for $\sigma_-$ and
$E=2t \sqrt{1+cos^2(k_0+\pi/2)}$ for $\sigma_+$. As the comb's size gets bigger, almost all the states- i.e., all the
states apart from a set of measure zero - belong to $\sigma_0$; in fact, since
$\int_{\sigma_0}\rho_0(E)dE=L_1\cdot (L_2-1)$, the normalized
density of states belonging to $\sigma_0$ tends to 1 while, for what
concerns the hidden spectrum, one has $\lim_{L_1,L_2\to
\infty}\int_{\sigma_{\pm}} \rho_\pm(E)\,dE=0$. Since the hidden spectrum does not contribute to the
normalized density of states $\rho(E)$ of the pure hopping model on the comb,
one has:
\begin{equation}
\rho (E) =\frac{1}{\pi \sqrt{4t^2-E^2}}. \label{rho_norm_chain}
\end{equation}
Normalizing the density of states of the lower hidden spectrum to $L_1$, one
obtains for $E\in \sigma_-$:
\begin{equation}
\frac{1}{L_1}\, \rho_-(E)=\frac{|E|}{2\pi
\sqrt{8t^2-E^2}\sqrt{E^2-4t^2}} \label{den2}
\end{equation}
An analogous equation holds for the spectral region $E \in \sigma_+$. The density of states
can then be plotted as in Fig.~\ref{combo2}, where the
pertinent normalizations for the continuous and hidden part of the spectrum have been used.

The lowest energy eigenvalue in the normalized density of states is $E_d=-2t$;
but, this is not the lowest energy attainable by a particle hopping
on a comb since there is a lowest localized eigenstate belonging to
$\sigma_-$ whose eigenvakue is given by:
\begin{equation}
E_{0}=-2 \sqrt{2}t. \label{gs_en_comb}
\end{equation}
In fact, for $L_2 \to \infty$, since the energy is a decreasing
function of $h$, the lowest energy level of $\sigma_-$ is attained when $\cos(k_0)=1$.

Of course, $E_0<E_d$ and this should indicate that the spectrum is gapped.
However, one does not find an energy gap between
$E_0$ and $E_d$, since, for each value of $k_0$ ($\cos(k_0)>0$) there is a solution of Eq.
\eqref{low_hs} with a different energy in the interval $[E_0,E_d]$. In a finite comb of $L_1 \times L_2$ sites there are $L_1/2$
solution of this type and, as $L_1\to \infty$, these solutions fill
densely the interval $[E_0, E_d]$. These spectral proprieties should be contrasted with the ones arising when non
interacting quantum particles are trapped in a harmonic well. 

\begin{figure}
\begin{center}
\includegraphics[scale=0.4]{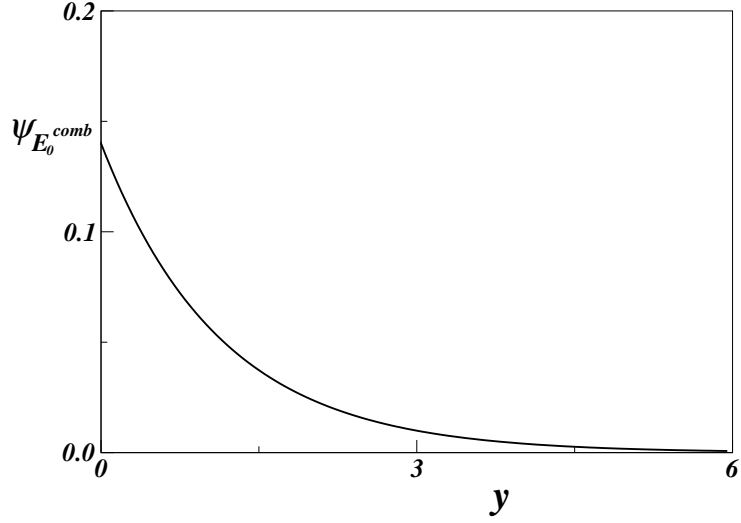}
\caption{\label{combo1} The normalized ground-state wavefunction
of the comb network as a function of the distance $y$ from the
origin for a comb with $51 \times 51$ sites.  }
\end{center}
\vspace{8mm}
\end{figure}

From Eqs. \eqref{sol2} and \eqref{eq_del} it is also possible to
show that, when $L_1, L_2 \to \infty$, the
eigenvector corresponding to the lowest energy eigenvalue is
\begin{equation}
\label{ground_state_comb_2}
\psi_{E_0}(x,y)=\frac{1}{2^{1/4}}e^{- \vert y \vert /\xi} ,
\end{equation}
where $\xi=1/h=1/\log(1+\sqrt{2})$ is the parameter accounting for
the  localization around the backbone. In Fig.~\ref{combo1} we plot
the ground-state wavefunction as a function of the distance from the
backbone; the plot well evidences the exponential localization only around the comb's backbone.

Although not explicitly imposed, the condition of the continuity
of the discrete gradient holds at the points $(0,y)$. Of course one
cannot ask for the continuity of the linear derivatives along $x$ or
$y$ since it is just this discontinuity which allows a particle
moving on the finger to hop in the direction of the backbone.

\begin{figure}
\begin{center}
\includegraphics[scale=0.4]{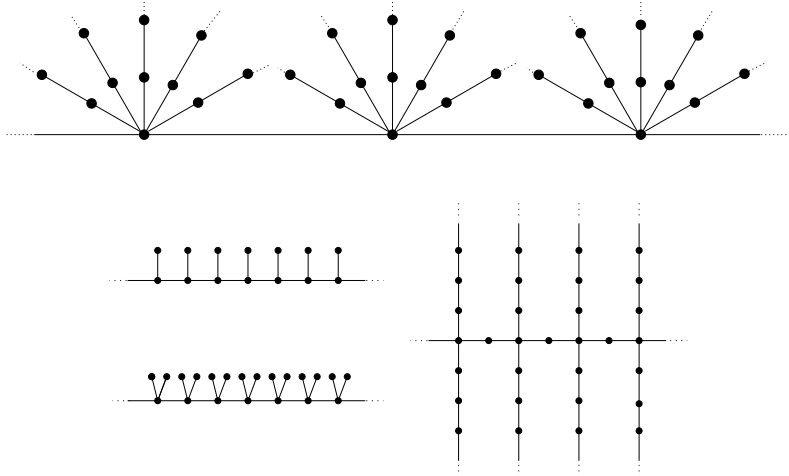}
\caption{\label{combo3}Top: The star-comb network.  Bottom Left: Two
``mini''-comb networks (top: $p=1$; bottom: $p=2$). Bottom Right:
the ``semi''-comb network. }
\end{center}
\vspace{8mm}
\end{figure}

A similar analysis may be carried to determine the spectrum of  a quantum particle hopping on different comb-like
networks, such as the star-comb, mini-comb, and semi-comb
depicted in Fig.~\ref{combo3}.  It is not difficult to convince
one-self that, for a pertinent choice of boundary conditions, the
adjacency matrix describing the connectivity of these bundled graphs
admits also an hidden spectrum.

In the following we shall report only the lowest energy eigenvalues
pertaining to quantum particles hopping on these networks. For  a star-comb \cite{mancini06}
one has
\begin{equation}
E_0^{sc}=-t\, \frac{(p- 2)+ p\,  \sqrt{p}}{p -1} \,,
\label{ground_state_en_sc}
\end{equation}
 where $p$ is the number of arms on each star.  For the semi-comb, one finds
\begin{equation}
E_0^{semi} = - 2 \sqrt{\phi} \, t, \label{e0_semi}
\end{equation}
where $\phi=(1+\sqrt{5})/2 \simeq 1.618$ is the golden section. For
a mini-comb, one finds
\begin{equation}
E_0^{mini}=- \left( 1+\sqrt{1+p}\right) t. \label{e0_mini}
\end{equation}

As we shall see in the following, the value of the ground-state
energy is intimately tied to the critical temperature at which
quantum bosons hopping on comb-networks undergo a
spatial Bose-Einstein condensation.

\section{Complex Behaviors emerging from the network's connectivity}

In this section we review how the pertinent choice of the network's connectivity leads to the emergence of new phenomena
in quantum devices realized with superconducting JNNs \cite{sodano06} and cold atoms in optical lattices \cite{giusiano04, burioni00}.
Our subsequent analysis well evidences that new emerging phenomena
are possible only if the network's connectivity is described by an adjacency matrix supporting an hidden spectrum. As we shall see, the states belonging
to the hidden spectrum induce in a discrete many body system the onset of a new relevant energy scale and enhance the number of states which can be occupied by 
quantum particles at low energy. 

\subsection{Enhanced Josephson critical currents in a comb-shaped JJN}

In the following, we use the LBdG equations derived in Section 2 to compute the zero-voltage Josephson
critical currents of junctions located on Josephson
linear chains and comb-shaped Josephson networks. Using the
eigenfunctions of the LBdG equations, a self-consistent computation
yields for both systems the gap function, the chemical potential and
the quasi particle spectrum. Our analysis evidences that, on the backbone of a comb-shaped JJN,
the BCS equations are satisfied with a renormalized value of the
Josephson energy. Then, we compute the zero-voltage Josephson
critical currents $I_c$ on the comb's backbone and compare our
results for $I_c$ with the outcomes of the experimental measurements presented in \cite{silvestrini05,sodano06} and
summarized in Fig.5.

For a comb network with $L \times L$ islands (see Fig.1), one finds
a solution of the LBdG equations (\ref{LBdG1}) and (\ref{LBdG2}) where
both the Hartree-Fock potential $U(i)$ and the gap function
$\Delta(i)$ are position dependent. The eigenvalue
equation
\begin{equation}
-E_J\sum_j A_{ij} \psi_\alpha(j) = e_\alpha
\psi_\alpha(i), \label{spect}
\end{equation}
leads to the emergence of the hidden spectrum analyzed in section 3.

For a crude analytical estimate, one may
require that, away from the backbone, the fingers may be regarded as
a linear chain with uniform potentials (i.e., $\Delta(i)=\Delta_c$
and $U(i)=U_c$). To get coupled equations for $\Delta_b$,
$\Delta_c$, $U_b$, and $U_c$, one writes the LBdG equations
(\ref{LBdG1}) and (\ref{LBdG2}) on a generic backbone's grain $i$. Upon setting
$u_\alpha(i)=U_\alpha \psi_\alpha(i)$ and $v_\alpha(i)=V_\alpha
\psi_\alpha(i)$, with $U_\alpha^2+ V_\alpha^2=1$, the
self-consistency equation for $U$ implies that, at $T=0$, $U_b
\approx U_c - \frac{\tilde{\cal V} C_0^2}{2}$; upon requiring
$\tilde{\mu} \approx U_b $ one immediately sees that, due to the
localized modes of the hidden spectrum, the chemical potential on
the comb's backbone is smaller than the one measured on the chain.

By substituting the wavefunctions of the eigenstates of the hidden
spectrum \cite{burioni00} in Eqs.~(\ref{LBdG1}) and (\ref{LBdG2}) and
using $ \tilde{\mu} \approx U_b $ one gets
\begin{equation}
\Delta_b=\Delta_c+ \frac{\Delta_b \tilde{\cal V}}{\pi} \cdot
\int_{0}^{\pi/2} dk \frac{\cos{k}}{\in_k \sqrt{1+\cos^2{k}}} \cdot
\tanh{ \left( \frac{\beta}{2} \in_k \right) }. \label{D-U-1}
\end{equation}
where $\in_k=\sqrt{\Delta_b^2 + 4 E_J^2 \left( 1+\cos^2{k}
\right)}$. The hidden spectrum eigenstates contribute also to the gap
function $\Delta_b$ through the second term in the rhs of
Eq.~(\ref{D-U-1}): without an hidden spectrum, $\Delta_b$ equals $\Delta_c$.

When $E_J \gg \Delta_b, \Delta_c$, Eq.~(\ref{D-U-1}), at $T=0$,
yields
\begin{equation}
\frac{\Delta_b(T=0)}{\Delta_c(T=0)}=
\frac{1}{1-\frac{\eta_c \tilde{\cal V}}{2 \pi E_J} } \equiv {\cal
K} \label{rip1}
\end{equation}

where $\eta_c \equiv (1/\sqrt{2}) \, \log{ \left( 1+\sqrt{2}
\right) }$. Furthermore, at low temperatures,
\begin{equation}
\Delta_b(T) /
\Delta_c(T) \approx \Delta_b(T=0) / \Delta_c(T=0). \label{rip2}
\end{equation}
Using the
parameters $E_J$ and $\tilde{\cal V}$ obtained from the measurements
carried on the JJ chain (see Appendix A), one gets ${\cal K} \approx
1.13$.

Upon requiring that the $T=0$ backbone's
gap function has a BCS like functional form- i.e., $\Delta_b(T=0)=8
\bar{E}_J e^{ -2 \pi \bar{E}_J / \bar{ {\tilde{\cal V}} } }$, with
$\bar{E}_J$ and $\bar{ {\tilde{\cal V}} }$ being the renormalized
Josephson energy and the renormalized interaction term- one is able
to estimate the renormalization of the Josephson coupling within the
LBdG approach. Namely, one has
\begin{equation}
\bar{E}_J={\cal K} E_J; \, \, \, \, \, \, \, \, \, \bar{
{\tilde{\cal V}} }={\cal K} {\tilde{\cal V}}, \label{ren}
\end{equation}
which embodies the effects of the hidden spectrum on the Josephson
current.
In Fig. 5 we plot, as a function of the normalized temperature, the
values of $I_c$ measured in \cite{silvestrini05}
(squares) and the values of $I_c$ obtained from the
Ambegaokar-Baratoff \cite{amba} formula using both the renormalized coupling
given by Eq.~(\ref{ren}) and the gap function along the backbone for
the comb-like JJN studied in \cite{silvestrini05,sodano06} (solid curve): the
agreement of the results of the LBDG analysis with the outcomes of experiments is very good at low
temperature.
\begin{figure}
\begin{center}
\includegraphics[scale=.45, angle=270]{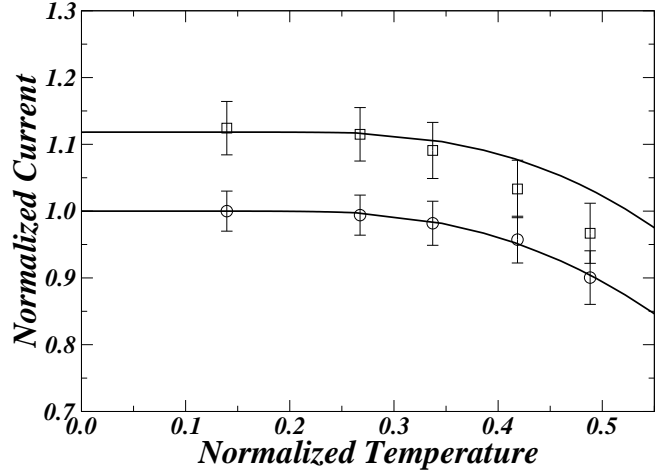}
\caption{\label{ambbar} Critical currents (in units of the critical current on the
reference chain at $T=1.2 K$) as a function of $T/T_c$ for the
backbone and the chain. The solid lines are the estimated critical
currents for the backbone (top) and the chain (bottom). Circles
(squares): experimental values for the chain (backbone).}
\end{center}
\vspace{8mm}
\end{figure}

\subsection{Spatial BEC of bosons hopping on a comb}

The thermodynamic properties of non-interacting bosons hopping on a
comb evidences the emergence of a spatial BEC on the backbone even
if the euclidean dimension of a comb is $1$
\cite{burioni00,burioni01}. In this section we shall review the
argument showing how this emergent behavior follows from the
existence of an hidden part in the spectrum of the adjacency matrix
describing the comb's connectivity; in fact, the presence of a dense set of states filling the gap between
$-2t$ and $E_0$ induces a change in the chemical
potential, which, in turn, allows for the existence of a finite value of the 
critical temperature at which spatial BEC occurs.

Fixing the number of particles in the grand canonical partition
function amounts to choose the fugacity $z$ so that
\begin{equation}
N_T=\sum_{E\in \sigma}
\frac{d(E)}{z^{-1} e^{\beta (E-E_0)}-1}.
\label{filling1}
\end{equation}
In Eq.~(\ref{filling1}) $N_T$ is the number of particles, $d(E)$ is
the degeneracy of each single-particle eigenstate, $\beta=1/k_B T$
and $E_0$ is the energy of the lowest energy state allowed to a
particle hopping on a comb; the sum is taken over the entire
spectrum $\sigma$.

For free bosons hopping on a square comb of size $L$ (with $N_S=L \times L$
sites), one has
\begin{equation}
N_T=N_{E_0}(L,T)+N_{\sigma_{-}}(L,T)
+N_{\sigma_{+}}(L,T)+
\int_{E \in \sigma_0} \, dE \,
\frac{L^2 \rho(E)}{z^{-1} e^{\beta (E+\sqrt{8}t)}-1 },
\label{filling}
\end{equation}
where $N_{E_0}(L,T)$, $N_{\sigma_{-}}(L,T)$ and
$N_{\sigma_{+}}(L,T)$ denote, respectively, the number of particles
which, at a certain temperature $T$, occupy the ground-state and the
two regions $\sigma_{-}$ and $\sigma_{+}$ of the hidden spectrum;
$\rho(E)$, with $E\in \sigma_0$, is the energy density of states
defined in Eq. \eqref{rho_catena}. Upon defining $n_{E_0}=N_{E_0}/L^2$, 
$n_{\sigma_0}=N_{\sigma_0}/L^2$ and $n_{\sigma_{\pm}}=N_{\sigma_{\pm}}/L^2$
as the contribution given by each one of the three spectral regions to the number of
particles per site, one has - as $L \to \infty$ - that
\begin{displaymath}
n_{E_0}(T) = \lim_{L \to \infty}\frac{1}{N_S}
\frac{1}{z^{-1}-1},\label{uno}
\end{displaymath}

\begin{displaymath}
n_{\sigma_{-}}(T)= \lim_{L\to \infty} \frac{2}{L^2}
\sum_{n=1}^{(L-1)/4} \frac{1}{z^{-1} e^{\beta t
[\sqrt{8}-2\sqrt{1+\cos^2(2\pi n/L)}]}-1},\label{due}
\end{displaymath}
and
\begin{eqnarray*}
n_{\sigma_{+}}(T) &=& \lim_{L\to \infty} \frac{1}{L^2}
\sum_{n=1}^{(L-1)/4} \frac{2}{z^{-1} e^{\beta t
[\sqrt{8}+2\sqrt{1+\cos^2(2 \pi n/L+\pi/2)}]}-1}  \\
&<& \lim_{L\to \infty}\frac{2}{L}\: \frac{1}{z^{-1} e^{\beta t(
\sqrt{8}+2)}-1 }=0 \quad  \quad \forall \: T. \label{n_sigma}
\end{eqnarray*}
The last equation shows that, at any finite temperature $T$,
$\sigma_{+}$ is not macroscopically occupied.

The last term of the right-hand side of Eq. \eqref{filling} is the
number of bosons in the delocalized states. On a chain the integral
appearing in Eq. \eqref{filling} is diverging since the limit $z \to
1$ is attained when the energy equals $-2t$; at variance, for a
comb, the existence of the hidden spectrum renders the same integral
convergent since the limit $z \to 1$ is now attained at the lower energy
$- 2 t \sqrt{2}$ which lies outside the interval $-2t,2t$. One then
sees explicitly  how the network's inhomogeneity works to induce the emergence
of a spatial BEC even if the bosons are "free" and the network is
one-dimensional.

If one defines as $T_c$ the critical temperature at which spatial
BEC on the comb's backbone occurs, then, for any  $T<T_c$, the
localized ground-state is macroscopically filled. Since
$n_{E_0}(T_c)=0 $, using Eqs.~(\ref{rho_catena}) and (\ref{filling}),
 $T_c$ may be determined (as a function of the filling fraction $f$ and
 of the hopping strength $t$) from
\begin{equation}
\pi f=\int_{-2t}^{2t} \frac{dE}{\sqrt{{4t^2-E^2}} }
\frac{1}{e^{(E-E_0)/(k_{B}T_{c})}-1 }. \label{critical_T}
\end{equation}
Equation (\ref{critical_T}) can be solved numerically for any
value of $f$. When $f \gg 1$, one may expand the exponential in
Eq. \eqref{critical_T} to the first order in the inverse of the
critical temperature $T_c$ getting
\begin{equation}
\frac{\pi f}{k_B T_c} \approx \int_{-2t}^{2t} dE
\frac{1}{\sqrt{4t^2-E^2}} \frac{1}{E-E_0}.
\label{critical_t2}
\end{equation}
Substituting $\cos{\theta}=E/2t$ in Eq.~(\ref{critical_t2}), one has
\begin{displaymath}
\frac{2 t \pi f}{k_B T_c}=
\int_{0}^{\pi} \frac{d\theta}{\cos{\theta}-E_0/2t},
\end{displaymath}
from which
\begin{equation}
k_B T_c =E_J \sqrt{\Bigg( \frac{E_0}{2t} \Bigg)^2-1},
\label{ris_temp_crit}
\end{equation}
with $E_J$ being the {\em Josephson energy} defined in Eq.~(\ref{Jos-ij}).
Equation (\ref{ris_temp_crit}) allows for a simple estimate of the critical temperature $T_c$
at which BEC occurs for free bosons hopping on a comb. Upon inserting the pertinent value 
of the ground-state energy $E_0$
in Eq. \eqref{critical_t2}, one finally gets
\begin{equation}
k_B T_c \approx  E_J. 
\label{t_c_f_comb}
\end{equation} 
A similar analysis can be carried out for the variety of comb-shaped networks depicted in fig. 4.

The condensate's fraction may be easily determined as a
function of the scaled temperature $\tau=\frac{T}{T_c}$. Taking into account that
\begin{equation}
N_{\sigma_{0}}(\tau)= \lim_{L \to \infty} N_S \int_{-2t}^{2t}
\rho(E) \frac{dE}{e^{\beta (E-E_0)}-1 }\approx N_T \cdot \tau,
\label{n_B}
\end{equation}
expanding to the first order in $\beta$ the exponential appearing in Eq.~(\ref{n_B}) \cite{nota}, 
and, finally, using Eqs.~(\ref{filling}) and (\ref{n_B}), one may easily compute the number of
particles occupying the states belonging to $\sigma_-$. Namely, one has to compute 
$N_0=N_{E_0}+N_{\sigma_-}$, where $N_{E_0}$ ($N_{\sigma_-}$) is the number of particles 
in the ground state (in $\sigma_-$, except for those occupying the ground state): as a result the
fraction of condensate, for $T<T_c$, is given by
\begin{equation}
\frac{N_{0}}{N_T} \approx 1 - \tau . \label{n_0}
\end{equation}
For $f$ ranging from $10^3$ to $10^9$, the results provided by Eq.~(\ref{n_0}) 
differ from those obtained by the numerical evaluation
of $N_{0}$ from Eq. \eqref{filling} by less than $1 \%$.
Equation (\ref{n_0}) shows that the condensate has dimension $1$.

Finally, one may compute the average number of bosons
$N_{B}(x,y)$ occupying a generic site $(x,y)$ of a comb \cite{giusiano04}; of course, this number depends only on $y$ due to the
translational invariance of a comb along the backbone.
One finds that $N_{B}(y;\tau)$ is given by:
\begin{equation}
\begin{split}
N_{B}\left(y;\tau \right) &=
N_{E_0}\left(\tau\right)\: \vert \psi_{E_0}(y)\vert^2+
\sum_{E_{n}\in \sigma_{-}}N_{\sigma_{-}}\left(E_n;\tau \right)
\:\vert \psi_{E_{n}}(y) \vert^2
\\
&+ L^{2}
\int_{E \in \sigma_0} dE \:\rho(E) \:
\frac{1}{e^{\beta (E+\sqrt{8}t)}-1} \:
\vert \psi_{E}(y)
\vert^2.
\end{split}
\label{localizzazione}
\end{equation}
In Eq.~(\ref{localizzazione})
$\psi_{E_0}(y)$ is the wavefunction
corresponding to the ground-state of the single-particle spectrum and
$\psi_{E_n}(y)$ are the eigenfunctions corresponding to the energies
$E_n$ of the hidden spectrum $\sigma_{-}$; $N_{\sigma_{-}}
(E_n)$ is the number of particles with energies $E_n \in \sigma_{-}$
and $N_{E_0}$ is the number of particles in the ground-state.
In the last term of Eq.~(\ref{localizzazione}) $\psi_E(y)$ are the
delocalized eigenfunctions of the eigenvalue equation (\ref{eigen_eq}).
For determining $N_{B}(y;\tau)$, one needs to compute
$N_{E_0}$ and $N_{\sigma_{-}}$, which are evaluated \cite{giusiano04} in Appendix B.
Using these results, an explicit analytical form for the number of bosons at site $y$, $N_B(y)$,
may, then, be derived.
The last term in Eq.~(\ref{localizzazione}) yields, in fact,
the contribution coming from the delocalized states:
for a large network ($L \gg 1$), and far away from the backbone, this number
is independent on the site index $y$ and equals
a constant $(N_T/L^2) \tau$.  Using Eqs.~(\ref{n_0_corr}) and (\ref{st_nasc_ap}),
for $\tau<1$, one finds that the distribution of bosons is, for $y \gg 1$, given by \cite{giusiano04}
\begin{equation}
\label{rapporto_sempl}
\frac{N_B (y;T/T_c)}{f} \approx \tau .
\end{equation}

\begin{figure}[t]
\begin{center}
\includegraphics[scale=.4,angle=270]{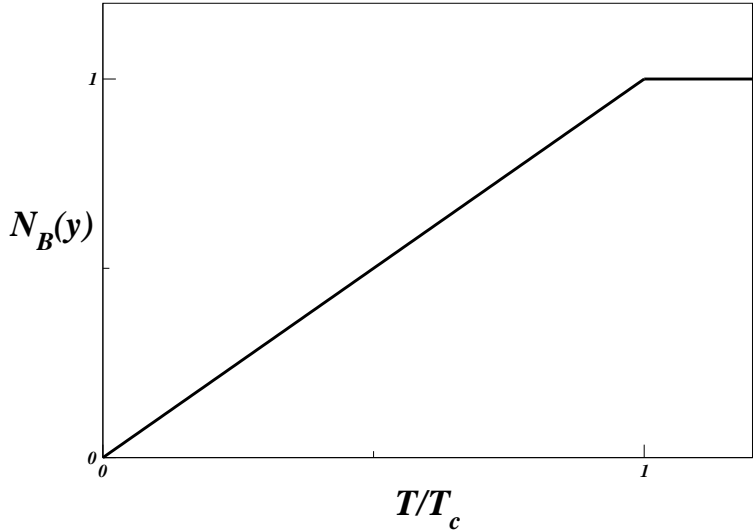}
\caption{\label{fig5}
Distribution of the number of bosons $N_B$ as a function of
$T/T_c$ computed for $y \gg 1$, very far from the backbone.
$N_{B}(y)$ is in units of the
filling $f$ and is therefore equal to 1 for $T\geq T_c$.
}
\end{center}
\vspace{8mm}
\end{figure}

The signature of the emerging spatial BEC in a system of non-interacting bosons
hopping on a comb-shaped network is provided then by the remarkably sharp
decrease of the number of bosons at sites located away from the
backbone. The linear dependence exhibited by the solid
line in Fig.~\ref{fig5} is consistent with the observation that the condensate has dimension 1.

\section{Concluding Remarks}
We evidenced how the optimal engineering of the {\em shape} of a network
leads to the emergence of complex features in quantum devices realized with bosonic and
superconducting networks, whose connectivity is described by an adjacency matrix supporting an hidden spectrum. 
For this purpose we analyzed the paradigmatic case of comb-shaped bosonic and superconducting networks.

For free bosons hopping on a comb, we evidenced how the network's
connectivity is responsible for the emergence of a spatial BEC along
the comb's backbone and computed the critical
temperature where a spatial BEC emerges. We then analyzed the
inhomogeneous distribution of the bosons along the comb fingers and estimated the dependence of the 
non-condensate fraction on the reduced
temperature $\tau$; we showed that the signature for the emergence of a spatial BEC on the comb's backbone is 
provided by a rather sharp decrease of
the number of bosons occupying the fingers as the temperature is lowered below $T_c$.  Finite size corrections to 
our results are negligible already for
$f\sim 100$. With little modifications our analysis could be carried
out also for diverse network's topologies supporting BEC
\cite{burioni01}. 

We analyzed also superconducting JJNs fabricated on a comb-shaped insulating support. We showed that a non perturbative 
(i.e., induced by the states of the hidden spectrum) renormalization of
some (i.e., the ones located on the backbone) of the Josephson couplings of a comb-shaped JJN is responsible
for the observed enhancement of $I_c$ of the Josephson junctions
located along the comb's backbone. We used an effective theory based
on the BdG equations since it allows for a simple and rather
intuitive derivation of Eq.~(\ref{D-U-1}), which 
evidences the crucial role played by the hidden spectrum in
determining the enhancement of the Josephson current along the
comb's backbone. The BDG approach relies on a few key assumptions; namely, that the eigenfunctions
of the BdG equations may be written in a tight binding form and that
only the electrons close to the Fermi surface contribute to determine
$E_J$; once these assumptions are made, one is able to derive Eqs.~(\ref{LBdG1})- (\ref{LBdG2}) 
and to account for all the dependence
on the electronic states into the definition of the parameters $E_J$
and ${\tilde{\cal V}}$, which may be determined \cite{sodano06} from the
measurements carried  in \cite{silvestrini05} on the linear chain. Our approach yields a
value of the renormalized Josephson coupling of the junctions
located on the comb's backbone in excellent agreement with the
experimental results (see fig.5). Similar phenomena
happen for the class \cite{burioni01} of JJNs fabricated on graphs whose
adjacency matrix supports an hidden spectrum.

An alternative way to look at comb-shaped networks is to regard them as a linear chain immersed
in an environment mimicked by the addition of the fingers.  This situation may be analyzed using either the
Caldeira-Leggett \cite{cale} or - for superconducting devices- the electromagnetic environment
\cite{elen} models. For Josephson devices this point of view was advocated long ago in \cite{schmi}. 
For these devices one expects that the nominal value of the Josephson energy $E_J$ of the
junctions in the array gets renormalized by the interaction with the
environment. However, one usually assumes that the
{\em effective} boundary conditions for the quantum fluctuations of
the environment modes do not depend on the Josephson couplings or on
the network's topology: while this assumption may be perfectly
legitimate for weak environmental fluctuations, better care should
be used if these fluctuations are strong as it may well happen for
one dimensional JJNs. A simple paradigmatic example of a non
perturbative renormalization of Josephson couplings is given by the
simple inhomogeneous one-dimensional array analyzed in
\cite{glala,giuso}, where the source of inhomogeneity is given by
putting on a site of the linear chain a {\em test} junction with a
different nominal value of the Josephson coupling $E_J$. Our
analysis shows that, for a comb-shaped JJN, the
Josephson couplings on the backbone get renormalized and that this
renormalization is non perturbative since the peculiar connectivity
of a comb modifies the spectrum of quantum modes living on linear
chains by the (obviously non-perturbative) addition of an infinite
set of localized states below the continuum threshold: adding the
fingers to a backbone chain is, in fact, a topological {\em
operation} since it amounts to a non trivial change of boundary
conditions for the Josephson linear chain. It would be interesting to investigate in this perspective also systems realized with cold atoms 
living on pertinent inhomogeneous optical lattices.

Our analysis provides experimentally testable examples of how the space connectivity affects coherent behaviors of physical systems.

\section*{Acknowledgements}
We benefited from discussions with M. Cirillo, D. Giuliano, G. Giusiano, A. Montorsi, M. Rasetti, B. Ruggiero, R.
Russo, P. Silvestrini. Our research has been partially supported by the MUR Project 
{\it Josephson Networks for Quantum Coherence and
Information} (grant No.2004027555). 
P. S. thanks the {\em the Progetto Lagrange}  and the {\em Statistical Theory Group at S.I.S.S.A.} for financial support 
during his residence at the Politecnico di Torino and S.I.S.S.A..

\appendix

\section{Solution of LBdG for a JJ-chain}

Let us consider a Josephson chain with $N_S$ grains:
$i=1,\cdots,N_S$ and $\alpha \to k$, with $k=2\pi n/N_S$
and $n=-N_S / 2, \cdots, N_S/2 -1$ (periodic boundary conditions are
used). Since $A_{ij}=\delta_{i,j+1}+\delta_{i,j-1}$, one gets that the eigenvalues
of the matrix $\epsilon_{ij}$ are
\begin{equation}
E_k=-2t \cos{k}+U_c-\mu,
\label{E-chain}
\end{equation}
where $U(i) \equiv U_c$ does not depend on the grain position. Similarly,
one sets $\Delta(i) \equiv \Delta_c$.
One finds then $u_k(j)=N_S^{-1/2} U_k e^{ikj}$ and
$v_k(j)=N_S^{-1/2} V_k e^{ikj}$, which, in turn, lead to $U_k^2=(1/2)
[1+E_k/\in_k]$ and $V_k^2=(1/2)
[1-E_k/\in_k]$, where $\in_k=\sqrt{\Delta_c^2+E_k^2}$. From
(\ref{LBdG1}) one gets $1=(\tilde{\cal V}/2N_S) \sum_k \in_k^{-1}
\tanh{(\beta \in_k / 2)}$ and, for $N_S \gg 1$, one finally obtains
\begin{equation}
1=\frac{\tilde{V}}{4 \pi t} \int_{-2t}^{2t} \frac{dE}
{\sqrt{1-\frac{E^2}{4t^2}} \sqrt{\Delta_c^2+(E-\mu+U_c)^2}}
\tanh{\left( \frac{\beta}{2}  \sqrt{\Delta_c^2+(E-\mu+U_c)^2} \right)}.
\label{Delta-c-mu}
\end{equation}

A BCS-like behavior is obtained with $U_c-\mu \approx 0$:
Eq.~(\ref{Delta-c})
for $\Delta_c$ reads then
\begin{equation}
1=\frac{\tilde{V}}{4 \pi t} \int_{-2t}^{2t} \frac{dE}
{\sqrt{1-\frac{E^2}{4t^2}} \sqrt{\Delta_c^2+E^2}}
\tanh{\left( \frac{\beta}{2}  \sqrt{\Delta_c^2+E^2} \right)}.
\label{Delta-c}
\end{equation}
At first sight equation (\ref{Delta-c}) appears quite different
from the corresponding BCS gap equation (\ref{gapBCS}), basically
due to the factor $(1-E^2/4t^2)^{-1/2}$ appearing
in the integrand and originating from the density
of states on the chain. Nevertheless, the behavior is basically BCS.
In fact, at $T=0$, one gets
\begin{equation}
1=\frac{\tilde{V}}{2 \pi t} \cdot
\frac{1}{1+\frac{\Delta_c^2(T=0)}{4t^2}} \cdot
K\left( \frac{1}{1+\frac{\Delta_c^2(T=0)}{4t^2}} \right),
\label{Delta-c0}
\end{equation}
where $K(x)$ is the complete elliptic integral of first kind
\cite{abramowitz64}.
If one takes the limit $\Delta_c / t \ll 1$ (which is, in a sense,
the equivalent of the limit $n(0) V_{BCS} \ll 1$ in the BCS theory) and recalls that,
at $T=0$, $K(x) \approx (1/2) \log{16/(1-x)}$
\cite{abramowitz64}, one obtains
\begin{equation}
\Delta_c(T=0) \approx 8 t e^{-2 \pi t / \tilde{\cal V}},
\label{Delta-c0-r}
\end{equation}
which is the $1D$ equivalent on a chain of the well-known
BCS expression $\Delta(T=0)=2 \hbar \omega_D e^{-1/n(0) V_{BCS}}$.
At $T=T_c$ ($\Delta_c(T=T_c)=0$), one has
\begin{equation}
1=\frac{\tilde{V}}{2 \pi t} \int_0^1
\frac{d\epsilon}{\epsilon \sqrt{1-\epsilon^2}}
\tanh{ \left( \frac{\epsilon t}{2 k_B T_c} \right) },
\label{T-cc}
\end{equation}
which, for $\Delta_c / t \ll 1$, yields $1 \approx (\tilde{V}/2\pi t)
(\log{t/k_B T_c}+\log{{\cal C}})$ with ${\cal C} \approx 4.536$. As a result one gets 
\begin{equation}
k_B T_c={\cal C} t  e^{-2 \pi t / \tilde{\cal V}},
\label{T-cc-r}
\end{equation}
which is the $1D$ equivalent on a chain of the well-known
BCS expression $k_B T_c=1.14 \hbar \omega_D e^{-1/n(0) V_{BCS}}$.
Combining Eqs.~(\ref{Delta-c0-r}) and (\ref{T-cc-r}) enables to show that
\begin{equation}
\frac{\Delta_c(T=0)}{k_B T_c}=\frac{8}{{\cal C}} \approx 1.76
\label{rapp-c},
\end{equation}
which coincides with the result expected from the BCS theory.

\begin{figure}[t]
\begin{center}
\includegraphics[scale=.4, angle=270]{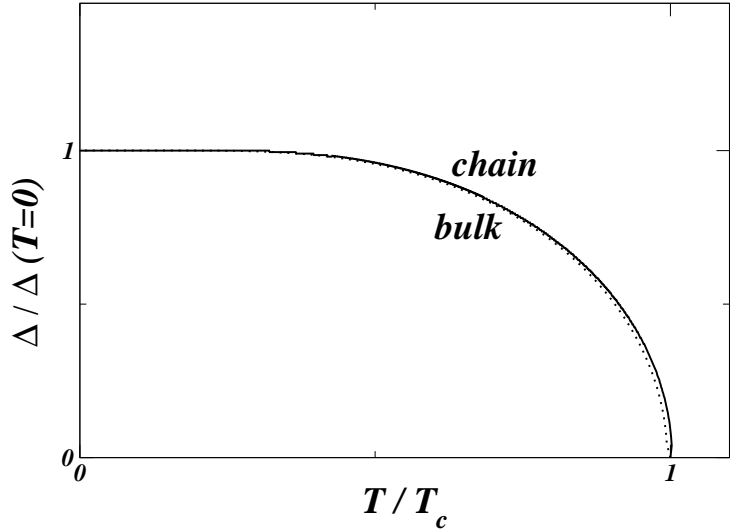}
\caption{\label{fig7}$\Delta/\Delta(T=0)$ vs.
$T/T_c$ obtained from the numerical solution of the BCS gap equation 
(\ref{gapBCS}) with the $Nb$ parameters (solid line) and
from the numerical solution of the gap equation for the
$1D$ chain of superconducting grains (\ref{Delta-c}), with the parameters
obtained for the setup of \protect\cite{silvestrini05} (dotted line).}
\end{center}
\vspace{8mm}
\end{figure}

For the experimental setup of \cite{silvestrini05,sodano06} one can estimate
$t \approx k_B \cdot 70 K$ and $\tilde{\cal V} / t \approx 1.75$, yielding
$T_c \approx 8.6 K$ and $\Delta_c(T=0) \approx k_B \cdot 15.9 K$.
In Fig.~\ref{fig7}, we plot
the ratio $\Delta(T)/\Delta(T=0)$ vs. the reduced temperature
$T/T_c$ obtained from the numerical solution of the BCS gap equation 
(\ref{gapBCS}) with the $Nb$ parameters for the bulk
($n(0) V_{BCS}=0.28$ and $\hbar \omega_D=k_B \cdot 275 K$) and the one
obtained from the numerical solution of the gap equation for the
$1D$ chain of superconducting grains (\ref{Delta-c}), with the parameters
obtained for the setup of \cite{silvestrini05}. The comparison
evidences the close similarity between the two curves.

Measurements on a chain made with $Nb$ grains yield $T_c \approx 8.8
K$ and $\Delta_c(T=0) \approx 1.4 meV \approx k_B \cdot 15.9 K$;
furthermore, in the experimental setup described in \cite{silvestrini05,sodano06} it
is $I_c \approx 18 \mu A$. The parameters $E_J$ and $\tilde{{\cal
V}}$, determined from Eq.~(\ref{T-cc-r}) are then given by $E_J \approx k_B \cdot 430 K$ and
$\tilde{{\cal V}}/ E_J =1.185$. 

\section{Distribution of bosons in the hidden spectrum}

In the thermodynamic limit, for $\tau \leq 1$, $N_{\sigma_{-}}(E_n,\tau)$
is given by \cite{buonsante02}
\begin{equation}
N_{\sigma_{-}} ( E_n;\tau)= \lim_{L \to \infty} L^2
\frac{2}{\frac{2\sqrt{2}t}{k_B T}(\pi n)^2 +\frac{L^2}{N_{E_0}(\tau)}}
\label{st_nasc_ex}
\end{equation}
and it depends on the number of particles in the ground-state
$N_{E_0}$.
Since $k_B T_c \approx 2 tf$,
from Eq. \eqref{st_nasc_ex}, it follows that the number of
particles occupying the hidden spectrum is given by
\begin{equation}
N_{\sigma_{-}}(\tau)=
\sum_{n=1}^{\infty} N_{\sigma_{-}} (E_n)=
-N_{E_0}+N_{E_0}
\sqrt{\frac{\tau}{\sqrt{2}} \frac{N_T}{N_{E_0}}} \;
{\rm{coth}} \, \left[ \sqrt{\frac{\tau}{\sqrt{2}} \frac{N_T}{N_{E_0}}}
\:\right]   .
\label{st_nasc_ex_tot}
\end{equation}
As a consequence of the fact that $N_0=N_{\sigma_{-}}+N_{E_0}$, from  Eq.~(\ref{n_0}), one finds the way to determine
$N_{E_0}/N_T$ as a function only of the scaled temperature $\tau$:
\begin{equation}
N_T (1 - \tau )=
N_{E_0}
\sqrt{\frac{\tau}{\sqrt{2}} \frac{N_T}{N_{E_0}}} \;
{\rm{coth}} \,
\left[\sqrt{\frac{\tau}{\sqrt{2}}\frac{N_T}{N_{E_0}}}\:\right].
\label{n_e0}
\end{equation}
Solving Eq.~(\ref{n_e0}) and substituting back the value obtained
for $N_{E_0}$ in Eq.~(\ref{st_nasc_ex}) allows for an exact numerical
evaluation of Eq.~(\ref{localizzazione}).

\begin{figure}
\includegraphics[scale=.32]{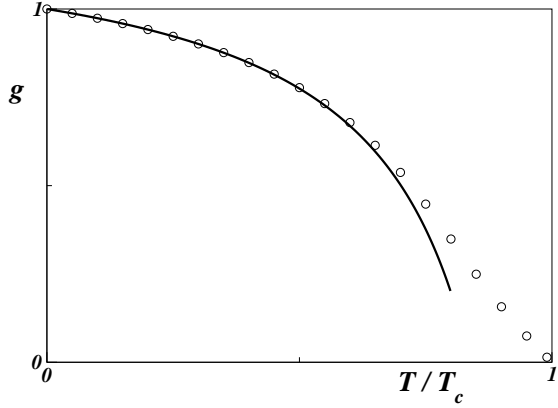}
\caption{ \label{confronto}
The function $g(\tau)$ defined in Eq.~(\ref{ansatz}): the empty
circles correspond to the numerical solution of the self-consistency
equation (\ref{self_cons1}); the solid line corresponds to the
approximate expression (\ref{self_cons2}).}
\vspace{8mm}
\end{figure}

Conventional wisdom supported by numerical evidence suggests that - apart from a small range of temperatures near $T_c$ - the largest
contribution to $N_0$ comes from $N_{E_0}$. Thus, it is physically
appealing to assume that $N_{E_0}$ is given by
\begin{equation}
N_{E_0}=N_T ( 1 - \tau ) g( \tau).
\label{ansatz}
\end{equation}
In Eq.~(\ref{ansatz}), $g(\tau)$ is a function only of the scaled
temperature $\tau$ and parametrizes the contributions to
$N_0$ coming from the states belonging to the hidden spectrum:
when $g=1$, the condensate is in the ground-state,
while, for $g=0$, is in the states of the hidden spectrum.
Substituting Eq.~(\ref{ansatz}) in Eq.~(\ref{st_nasc_ex}) and requiring
$N_0=N_{\sigma_{-}}+N_{E_0}$ with $N_0$ given by Eq.~(\ref{n_0}),
leads to a self-consistency equation for $g(\tau)$:
\begin{equation}
g(\tau) \sqrt{\frac{\tau}{\sqrt{2}(1-\tau) g(\tau)}} \;
{\rm{coth}} \, \left[\sqrt{\frac{\tau}{\sqrt{2}(1-\tau) g(\tau)}}
\:\right]=1. \label{self_cons1}
\end{equation}
For $\tau$ not too close to $1$,
a rather simple approximate solution of
Eq.~(\ref{self_cons1}) is provided by
\begin{equation}
g (\tau) \approx 2 - \sqrt{\frac{\tau}{\sqrt{2}(1-\tau)}}
 \; {\rm{coth}} \, \left[\sqrt{\frac{\tau}{\sqrt{2}(1-\tau)}} \:\right].
\label{self_cons2}
\end{equation}
The error made in using Eq.~(\ref{self_cons2}) instead of
the exact solution of Eq.~(\ref{self_cons1}) is within few percents:
for $\tau \le 0.5$ the error is less than $1 \%$, while for $\tau=0.7$
is about $5 \%$. In Fig.~(\ref{confronto}) we plot the function
$g(\tau)$ as obtained from the numerical solution of the
self-consistency Eq.~(\ref{self_cons1}) and from the approximate
expression (\ref{self_cons2}).

Upon inserting  Eq.~(\ref{self_cons2}) in Eq.~(\ref{ansatz}), one has
\begin{equation}
\label{n_0_corr}
\frac{N_{E_0}(\tau)}{N_T} \approx
(1- \tau )  \Bigg
(2 - \sqrt{\frac{\tau}{\sqrt{2}(1-\tau)}}\;
{\rm{coth}} \, \left[\sqrt{\frac{\tau}{\sqrt{2}(1-\tau)}}\:\right]
\Bigg), \end{equation}
and, from Eq.~(\ref{st_nasc_ex}), one gets
\begin{equation}
\label{st_nasc_ap}
\frac{N_{\sigma_{-}} ( \tau)}{N_T} =
\sum_{n=1}^{\infty}
\frac{2}{\frac{\sqrt{2}(\pi n)^2}{\tau}+\frac{1}{(1-\tau) g(\tau)}} \approx
( 1- \tau ) \Bigg\{
\sqrt{\frac{\tau}{\sqrt{2} (1-\tau)}} \;
{\rm{coth}} \, \left[ \sqrt{\frac{\tau}
{\sqrt{2} (1-\tau)}}\: \right] - 1 \Bigg\}.
\end{equation}


\begin{thebibliography}{99}

\bibitem{complex} N. Goldenfeld and L. Kadanoff, Science {\bf 284},
87 (1999).

\bibitem{enhare} E. Dagotto, Science {\bf 309}, 257 (2005).

\bibitem{nematic} P. Xiong et al., Phys. Rev. Lett. {\bf 69}, 3220 (1992); E. Fradkin and S. A. Kivelson,
Phys. Rev. {\bf 59}, 8065 (1999); S. A. Kivelson, E. Fradkin and
V.J. Emery, Nature {\bf 393}, 550 (1998).

\bibitem{manganite} E. Dagotto, T. Hotta and A. Moreo, Phys. Rep.
{\bf 344}, 1 (2001); E. Dagotto, {\em Nanoscale Phase Separation and
Colossal Magnetoresistance}, Springer-Verlag (2001).

\bibitem{stripe} V.J. Emery, S.A. Kivelson and J.M. Tranquada,
Proc. Natl. Acad. Sci. U.S.A. {\bf 96}, 8814 (1999); J. Zaanen,
Nature {\bf 404}, 714 (2000) and references therein.

\bibitem{pseudogap}Ch. Renner et al., Phys. Rev. Lett. {\bf 80}, 149
(1998); M. Suzuki, T. Watanabe and A. Matsuda, Phys. Rev. Lett.
{82}, 5361 (1999); T. Ekino et al. J. Low Temp. Phys. {\bf 117}, 359
(1999).

\bibitem{kivefra} S.A. Kivelson and E. Fradkin, {\em Preprint} cond-mat/0507459.

\bibitem{Ovchinnikov} Yu.N. Ovchinnikov, S.A. Wolf and V.Z. Kresin,
Phys. Rev. {\bf B63}, 064524 (2001).

\bibitem{anderson} P.W. Anderson, Science {\bf 177}, 393 (1972).

\bibitem{moshe} see, for instance: {\em Coherence in Superconducting
Networks}, J.E. Mooij and G. Sch\"on eds., Physica {\bf B152},
pp.1-308, (1988); {\em Josephson Junction Arrays}, H.A. Cerdeira and
S.R. Shenoy eds., Physica {\bf B222}, pp.253-406, (1996); R. Fazio
and H. van der Zant, Phys. Rep. {\bf 355}, 235 (2001).

\bibitem{morsch06} O. Morsch and M. K. Oberthaler, Rev. Mod. Phys.
{\bf 78}, 179 (2006); F.S. Cataliotti {\em et al.}, Science {\bf
293}, 843 (2001); B.P. Anderson and M. Kasevich, Science, {\bf
282}, 1686 (1998).

\bibitem{simanek94} E. Sim\`anek, {\em Inhomogeneous Superconductors},
Oxford University Press, New York, 1994.

\bibitem{MSS} Y. Makhlin, G. Shoen and A. Shnirman, Rev. Mod. Phys.
{\bf 73}, 357 (2001).

\bibitem{inhom} P.G. de Gennes, C. R. Acad. Sci. Ser. {\bf B292}, 279
(1981); P.G. de Gennes, C. R. Acad. Sci. Ser. {\bf B292}, 9 (1981);
S. Alexander, Phys. Rev. {\bf B27}, 1541 (1983); H.J. Fink, A.
Lopez, and R. Maynard, Phys. Rev. {\bf B26}, 5237 (1982); R. Rammal,
T.C. Lubensky, and G. Toulouse, Phys. Rev. {\bf B27}, 2820 (1983).

\bibitem{deut} G. Deutscher and R. Rosembaum, Appl. Phys. Lett {\bf
27}, 366 (1975); G. Deutsher, I. Grave and S. Alexander, Phys. Rev.
Lett. {\bf 48}, 1497 (1982); G. Deutscher {\em et al.}, Phys. Rev.
{\bf B24}, 6464 (1981).

\bibitem{consup} J. Berger, J. Rubinstein eds. {\em Connectivity and Superconductivity},
Lecture Notes in Physics, Springer-Verlag, Berlin, 2000.

\bibitem{ioffe} L.B. Ioffe {\em et al.}, Nature {\bf 415}, 503
(2002); B. Doucot, M.V. Feigel'man and L.B. Ioffe, Phys. Rev. Lett.
{\bf 90}, 107003, (2003); B. Doucot, L.B. Ioffe and J. Vidal, Phys.
Rev. {\bf B69}, 107003 (2003);  B. Doucot {\em et al.} Phys. Rev. B {\bf 71}, 024505
(2005).

\bibitem{wen} X.G. Wen and Q. Niu, Phys. Rev. {\bf B41}, 9377
(1990); X.G. Wen, Phys. Rev. Lett. {\bf 90}, 016803 (2003).

\bibitem{abilio} C.C. Abilio {\em et al.}, Phys. Rev. Lett. {\bf 83},
5102 (1999); J. Vidal, R. Mosseri and B. Doucot, Phys. Rev. Lett.
{\bf 81}, 5888 (1998).

\bibitem{feigel} I.V. Protopopov and M. V. Feigel'man, Phys.Rev.
{\bf B70}, 184519 (2004); I.V. Protopopov and M.V. Feigel'man,
cond-mat/0510766.

\bibitem{silvestrini05}  P. Silvestrini, R. Russo,
V. Corato, B. Ruggiero, C. Granata, S. Rombetto, M. Russo, M.
Cirillo, A. Trombettoni, and P. Sodano, cond-mat/0512478.

\bibitem{sodano06} P. Sodano, A. Trombettoni, P. Silvestrini, R. Russo,
and B. Ruggiero, New J. Phys., {\bf 8}, 327 (2006).

\bibitem{pitaevskii03}
L. P. Pitaevskii and S. Stringari, {\it Bose-Einstein Condensation},
Oxford University Press, Oxford, 2003.

\bibitem{oberthaler02} M. K. Oberthaler and T. Pfau,
J. Phys.: Condens. Matter {\bf 15}, R233 (2003).

\bibitem{roth03}
R. Roth and K. Burnett, Phys. Rev. {A 68}, 023604 (2003).

\bibitem{brunelli04}
I. Brunelli, G. Giusiano, F. P. Mancini, P. Sodano, and A.
Trombettoni, J. Phys. {\bf B 37}, S275 (2004).

\bibitem{oberthaler05}  M. Albiez, R. Gati, J. Folling,
S. Hunsmann, M. Cristiani, and M. K. Oberthaler, Phys. Rev. Lett.
{\bf 95}, 010402 (2005).

\bibitem{smerzi97} A. Smerzi, S. Fantoni, S. Giovanazzi, and S. R.
Shenoy, Phys. Rev. Lett. {\bf 79}, 4950 (1997).


\bibitem{graf} F. Harary, {\em Graph Theory}, Addison-Wesley,
Reading (1969).

\bibitem{burioni00}
R. Burioni, D. Cassi, I. Meccoli, M. Rasetti, S. Regina, P. Sodano,
and A. Vezzani, Europhys. Lett. {\bf 52}, 251 (2000).

\bibitem{burioni01} R. Burioni, D. Cassi, M. Rasetti, P. Sodano, and
A. Vezzani, J. Phys. B {\bf 34}, 4697 (2001).

\bibitem{giusiano04}
G. Giusiano, F. P. Mancini, P. Sodano, and
A. Trombettoni, Int. J. Mod. Phys. B  {\bf 18}, 691 (2004).

\bibitem{degennes} P.G. de Gennes, {\em Superconductivity of Metals
and Alloys}, Addison-Wesley (1989).

\bibitem{abeles} B. Abeles, Phys. Rev. B {\bf 15}, 2828 (1997).

\bibitem{jaksch98} D. Jaksch, C. Bruder, J. I. Cirac, C. W. Gardiner, and P. Zoller, Phys. Rev.
Lett. {\bf 81}, 3108 (1998).

\bibitem{mancini06}
F. P. Mancini, P. Sodano, and A. Trombettoni, to appear in ``Eleventh Training Course in the
Physics of Correlated Electron Systems and High-Tc Superconductors", Vietri sul Mare, Italy,
Oct 2006, cond-mat/0612388.

\bibitem{amba} V. Ambegaokar and A. Baratoff, Phys. Rev. Lett. {\bf
10}, 486 (1963); ibid. {\bf 11}, 104 (1963)

\bibitem{nota} This approximation holds for $f \gg 1$ and it is
in very good agreement with the results of a numerical evaluation of
the integral (\ref{n_B}).

\bibitem{cale} A.O. Caldeira and A.J. Leggett, Ann.Phys. (N.Y.)
{\bf 149}, 374 (1983).

\bibitem{elen} M.H. Devoret {\em et al.}, Phys. Rev.Lett. {\bf 64},
1824 (1990); S.M. Girvin {\em et al.}, {\bf 64}, 3183 (1990); G.
Sch\"on and A.D. Zaikin, Phys. Rep. {\bf 198}, 237 (1990).

\bibitem{schmi} A. Schmid, J. Low Temp. Phys. {\bf 49}, 609 (1982).

\bibitem{glala} L.I. Glazman and A.I. Larkin,
Phys. Rev. Lett. {\bf 79}, 3736 (1997).

\bibitem{giuso} D. Giuliano and P. Sodano, Nucl. Phys. {\bf B711},
480 (2005).

\bibitem{buonsante02}
P. Buonsante, R. Burioni, D. Cassi, and A. Vezzani,
Phys. Rev. B {\bf 66}, 094207 (2002).

\bibitem{abramowitz64} M. Abramowitz and I. A. Stegun,
{\em Handbook of mathematical functions
with formulas, graphs, and mathematical tables},
National Bureau of Standards, Washington, D.C., 1964.



 

\end{thebibliography}
\end{document}